\documentclass{ws-procs9x6}
\usepackage{amsfonts}
\usepackage{epsfig}
\usepackage[dvips]{color}

\begin{document}

\title{Long range hadron density fluctuations at soft $p_T$ 
in Au+Au collisions at RHIC}
\author{Mikhail L. ~Kopytine }
\address{Department of Physics, Kent State University, USA}
\author{for the STAR Collaboration}
\maketitle
\abstracts{
Dynamic fluctuations in the local density of non-identified hadron
tracks reconstructed in the STAR TPC are studied using the discrete wavelet
transform power spectrum technique which involves mixed event reference
sample comparison. The two-dimensional event-by-event analysis is performed
in pseudo-rapidity $\eta$  and azimuthal angle $\phi$  in bins of
transverse momentum $p_T$. 
HIJING simulations indicate that jets and mini-jets
 result in characteristic signals, visible already at soft $p_T$, 
when the dynamic texture analysis is applied.
In this analysis,
the discrepancy between the experiment and the HIJING expectations for Au+Au
at $\sqrt{S_{NN}}=200$ GeV
is most prominent in the central collisions 
where we observe the long range fluctuations to be enhanced
at low $p_T$, and suppressed above $p_T = 0.6$ GeV
}

\section{Introduction}

The on-going RHIC program, motivated by an interest in the bulk
properties of strongly interacting matter under extreme conditions,
has already yielded a number of tantalizing results.
Deconfinement and chiral symmetry restoration\cite{Meyer-Ortmanns:1996ea}
are expected to take place
in collisions of ultra-relativistic nuclei.
Because these phase transitions are multiparticle phenomena, a
promising, albeit challenging, approach is the 
study of dynamics of large groups of final state particles. 
The dynamics
shows itself in the correlations and fluctuations (texture) on a variety
of distance scales in momentum space.

The multi-resolution dynamic texture approach 
(applied for the first time\cite{NA44} at SPS)
uses discrete wavelet
transform \cite{DWT}(DWT) to extract such information.
At the present stage, the information is extracted in a comprehensive way, 
without any built-in assumptions or filters.
Mixed events are used as a reference for comparison in search for
dynamic effects. 
Event generators are used to ``train intuition'' in recognizing 
manifestations of familiar physics (such as elliptic flow or jets)
in the analysis output, as well as to quantify sensitivity to the 
effects yet unidentified, such as critical fluctuations or clustering 
of new phase at hadronization.

\section{The STAR experiment}
The STAR Time Projection Chamber\cite{STAR_TPC}(TPC) is
mounted inside a solenoidal magnet. 
It   tracks charged particles within a large acceptance ($|\eta|<1.3$,
$0<\phi<2\pi$) and is well suited for event-by-event physics and
in-depth studies of event structure.
The data being reported are obtained during the 
second ($\sqrt{S_{NN}}=200$ GeV) year
of RHIC operation. 
The minimum bias trigger discriminates on 
a neutral spectator signal in the Zero
Degree Calorimeters\cite{ZDC}.
By adding a requirement of high charged multiplicity within $|\eta|<1$ 
from
the scintillating Central Trigger Barrel, one obtains the central trigger.
Vertex reconstruction is based on the TPC tracking.
Only high quality tracks found to pass within 3 cm of the event vertex are 
accepted for the texture analysis.

\section{Dynamic texture analysis procedure}
\label{DWT}
Discrete wavelets are a set of functions, each having a proper width,
or scale, and a proper location so that the function differs from 0
only within that width and around that location.
The set of possible scales and locations is discrete.
The DWT transforms the collision event in pseudo-rapidity $\eta$ and azimuthal
angle $\phi$  into a set of two-dimensional functions.
The basis functions are defined
in  the ($\eta$, $\phi$) space and are
 orthogonal with respect to scale and location.
We accumulate texture information  
by averaging the power spectra of many events.

\begin{figure}
\epsfxsize=9cm
\epsfbox{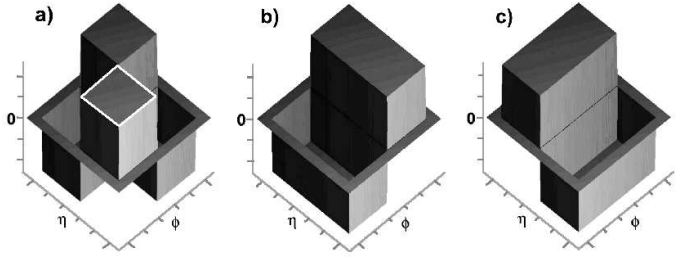}
\caption{
Haar wavelet basis in two dimensions. 
The three modes of directional sensitivity are:
a) diagonal b) azimuthal c) pseudo-rapidity. 
For the finest scale used, the white rectangle drawn ``on top'' of the 
function 
in panel a) would correspond to the smallest acceptance bin (pixel).
Every subsequent coarser scale is obtained by expanding
the functions of the previous scale by a 
factor of 2 in both dimensions. (Reproduced from {\protect\cite{NA44}}).
}
\label{haar}
\end{figure}
The simplest DWT basis is the Haar wavelet, built upon the \emph{scaling 
function}
$g(x) = 1$ for $0\le x<1$ and 0 otherwise.
The function
\begin{equation}
f(x) =     \{ +1 \mbox{\ for\ } 0\le x<\frac{1}{2}; 
                 -1 \mbox{\ for\ }  \frac{1}{2}\le x<1;
                  0 \mbox{\ otherwise}
              \}
\end{equation}
is the wavelet function.

The experimental acceptance in $\eta$,$\phi$, and $p_T$ 
(($|\eta|<1$, $0<\phi<2\pi$)) is
partitioned into bins. The $\eta$-$\phi$ partitions are of equal size,
whereas in $p_T$, the binning is exponential when more than one
$p_T$ bin is used.
In each bin, the number of 
reconstructed tracks satisfying the quality cuts is counted.

The scaling function of the Haar basis in
two dimensions (2D)
$G(\phi,\eta) = g(\phi)g(\eta)$
is just a bin's acceptance (modulo units).
The wavelet functions $F^{\lambda}$ 
(where the mode of directional sensitivity $\lambda$ can be 
azimuthal $\phi$, pseudo-rapidity $\eta$, or diagonal $\phi\eta$)
are
\begin{equation}
F^{\phi\eta}=f(\phi)f(\eta),\ \
F^\phi=f(\phi)g(\eta), \ \
F^\eta=g(\phi)f(\eta).
\end{equation}
We  set up a two dimensional (2D) wavelet basis:
\begin{equation}
F^{\lambda}_{m,i,j}(\phi,\eta) =
 2^{m}F^{\lambda}(2^{m}\phi-i,2^{m}\eta-j),
\label{wavelet_2D}
\end{equation}
where $m$ is 
the integer
scale fineness index,
$i$ and $j$ index the positions of bin centers in 
$\phi$ and $\eta$.
Then, $F^\lambda_{m,i,j}$ with integer $m$, $i$, and $j$ are known
\cite{DWT}
to form a complete orthonormal basis in the space
of all \emph{measurable functions} defined on the continuum of real
numbers $L^2({\mathbb{R}})$.
We construct $G_{m,i,j}(\phi,\eta)$ analogously to Eq.\ref{wavelet_2D}.

Fig. \ref{haar} shows the wavelet basis functions $F$ in two dimensions.
At first glance it might seem surprising that, unlike the 1D case, both $f$
and $g$ enter the wavelet basis in 2D.
Fig. \ref{haar} clarifies this: in order to fully encode an arbitrary
shape of a measurable 2D function, one considers it as an addition of a
change along $\phi$ ($f(\phi)g(\eta)$, panel (b)), 
a change along $\eta$ ($g(\phi)f(\eta)$, panel (c)), and
a saddle-point pattern ($f(\phi)f(\eta)$, panel (a)), 
added with appropriate weight (positive, negative or zero), for a variety
of scales.
The finest scale available is limited by the 
two track resolution, and,
due to the needs of event mixing, by the number of available events.
The coarser scales correspond to successively re-binning the track 
distribution.
The analysis is best visualized by considering the scaling function
$G_{m,i,j}(\phi,\eta)$ as binning the track distribution 
$\rho(\phi,\eta)$
in bins $i$,$j$
of fineness $m$, while the set of wavelet functions 
$F^{\lambda}_{m,i,j}(\phi,\eta)$ (or, to be exact, the wavelet expansion
 coefficients $\langle \rho, F^{\lambda}_{m,i,j}\rangle$)
gives the difference distribution between the data binned with given 
coarseness and that with binning one step finer.
We use WAILI\cite{WAILI} software to obtain the wavelet expansions.

In two dimensions, 
it is informative to present the three modes of a power
spectrum with different directions of sensitivity
$P^{\phi\eta}(m)$, $P^\phi(m)$, $P^\eta(m)$
separately.
We define the {\bf power spectrum} as
\begin{equation}
P^\lambda(m) =
\frac{1}{2^{2m}}\sum_{i,j}\langle \rho,F^\lambda_{m,i,j}\rangle^2 ,
\label{eq:P_m}
\end{equation}
where the denominator gives the meaning of spectral density
to the observable.
So defined, the $P^\lambda(m)$ of a random white noise field is
independent of $m$.
However, for physical events one finds  $P^\lambda(m)$ to be dependent
on $m$ due to the presence of {\bf static texture} features such as
acceptance asymmetries and imperfections (albeit minor in STAR),
and non-uniformity of the $\,dN/\,d\eta$ shape.
In order to extract the {\bf dynamic} signal, we use 
$P^\lambda(m)_{true}-P^\lambda(m)_{mix}$
where the latter denotes power spectrum obtained from the {\bf mixed events}.
The mixed events are composed of the ($\eta,\phi$) pixels of true events,
so that a pixel is an acceptance element of the finest scale used
in the analysis, and in no mixed event is there more than one pixel from 
any given true event.
The minimum granularity used in the analysis is $16\times16$ pixels.
\footnote{For a quick reference, here are the scales in $\eta$. Scale 1:
$\Delta\eta=1$; scale 2: $\Delta\eta=1/2$; scale 3: $\Delta\eta=1/4$ 
and so on.}

Systematic errors can be induced on
 $P^\lambda(m)_{true}-P^\lambda(m)_{mix}$ by the process of event
mixing.
For example, in events with different vertex position along the beam
axis, same values of $\eta$ may correspond to different parts of the TPC
with different tracking efficiency.
That will fake a dynamic texture effect in $\eta$.
In order to minimize such errors,
events are classified into {\bf event classes} with similar
multiplicity and vertex position.
Event mixing is done and  $P^\lambda(m)_{true}-P^\lambda(m)_{mix}$ is 
constructed within such classes.
Only events with $z$ vertex lying on the beam axis within 25 cm
from the center of the chamber are accepted for analysis.
To form event classes, this interval is further subdivided into five bins.
We also avoid mixing of events with largely different multiplicity.
Therefore, another dimension of the event class definition is that of
the multiplicity of high quality tracks in the TPC.
For central trigger events, the multiplicity range of an event
class is typically 50.

\section{``Coherent'' interference of patterns and normalization of
 power spectra}
\label{coherence}

Imagine a reconstructed event as a distribution of points in the space
of  variables $(\eta,\phi,p_T)$.  
We slice this space into $p_T$ bins and analyze two-dimensional 
($\eta,\phi$) patterns.
The patterns from different  $p_T$ slices of the  same event
will amplify the texture signal when those $p_T$ bins are merged.
Depending on how the amplification works, one will find different scaling
law to relate the $P^\lambda(m)_{true}-P^\lambda(m)_{mix}$
 signal amplitude  with the underlying
number of  particles.

The DWT power  spectrum at each  scale is (using
Haar wavelet) a sum of  squared pixel-to-pixel content differences for the
given  pixel fineness (scale).   One can  think of  the pixel-to-pixel
content difference the same way  as one thinks of a random fluctuation
in the pixel  content.  Imagine that the  pattern  being analyzed is a
small sub-sample  of the event,  and its number  of particles $N$  can be
increased at will, up to the point  of making it an entire event -- as is
the case when  the sub-sample is a $p_T$ bin of  the event.  
The pixel content  will scale with  $N$, and if the dynamic pattern 
preserves  its shape from one $p_T$ bin to another, 
the pixel-to-pixel
difference on the characteristic scale of the pattern will also scale as $N$.
Consequently,
the  dynamic component 
of the power spectrum for this scale will grow  as $N^2$.
We will call this behavior  ``coherent''
 in analogy  with optics, where  one needs
coherence  in order  to see  interference patterns.  

Normalization is needed in order to, first,
express different measurements in the same units; 
second, eliminate trends in $p_T$ dependence which 
are induced by the design of the measure and unrelated to the
physics.
For the ``coherent'' case,  the normalized dynamic texture observable is
\begin{equation}
(P^\lambda(m)_{true} - P^\lambda(m)_{mix})/P^\lambda(m)_{mix}/N.
\end{equation}
One  could also imagine   ``incoherent''  $p_T$   slices.   
In the  ``incoherent'' case,  the pixel
content will  grow proportionally to $N$, but  the pixel-to-pixel difference
will  grow as  the RMS fluctuation of  the pixel  content, i.e.  as the
Poissonian  $\sqrt N$.  
The dynamic component of the power spectrum will grow  as $N$ 
(i.e. $\propto P(m)$) and
\begin{equation}
(P^\lambda(m)_{true} - P^\lambda(m)_{mix})/P^\lambda(m)_{mix}
\end{equation}
should be used in this case.
In the DWT-based texture analysis, amplification of the
signal is based  not on adding the patterns  themselves, but on adding
the power spectra of local density fluctuations, that is (continuing
the optics analogy) adding the intensities rather than field amplitudes.
  For this reason, in the DWT analysis one
does not require ``coherence'' to  amplify the signals from many $p_T$
slices, just as in optics one does not need coherence to see the light
intensity  increase  with an  increase  in  the  number of photons.
 
\section{Textures of jets and critical fluctuations in event generators}
\label{jets_in_HIJING}
Dynamic  texture is  to be  expected  from HIJING\cite{HIJING}
  given its  particle
production  mechanism  at  RHIC  energy  (jets,  mini-jets  and  string
fragmentation).  
HIJING combines a perturbative QCD description at high $p_T$ with a model of
low $p_T$ processes.
\begin{figure}[tb]
\hspace{\fill}
\begin{minipage}[t]{52mm}
\epsfxsize=62mm
\epsfbox{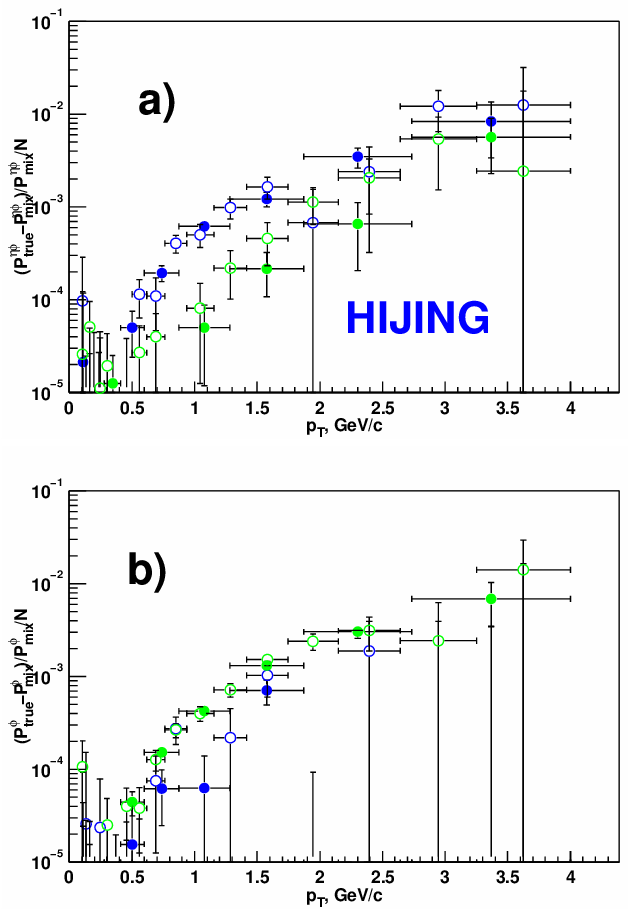}
\end{minipage}
\hspace{\fill}
\begin{minipage}[b]{52mm}
\epsfxsize=62mm
\caption{
a) -- $\eta\phi$, b) -- $\phi$ and c) -- $\eta$ directional components of the
dynamic texture in HIJING (events with impact parameter between 0 and 3 fm),
 arising primarily due to jets.
Data sets with different 
$p_T$ bin widths, indicated by the open and solid symbols,
 are statistically consistent at both scales when
 the ``coherent'' normalization is included.
\textcolor{blue}{$\bullet$} \textcolor{blue}{$\circ$}-- scale 1;
\textcolor{green}{$\bullet$} \textcolor{green}{$\circ$} -- scale 2.
Enhanced fineness scale 2 of  the $\phi$ texture plot (b)
reflects back-to-back correlations. 
}
\epsfbox{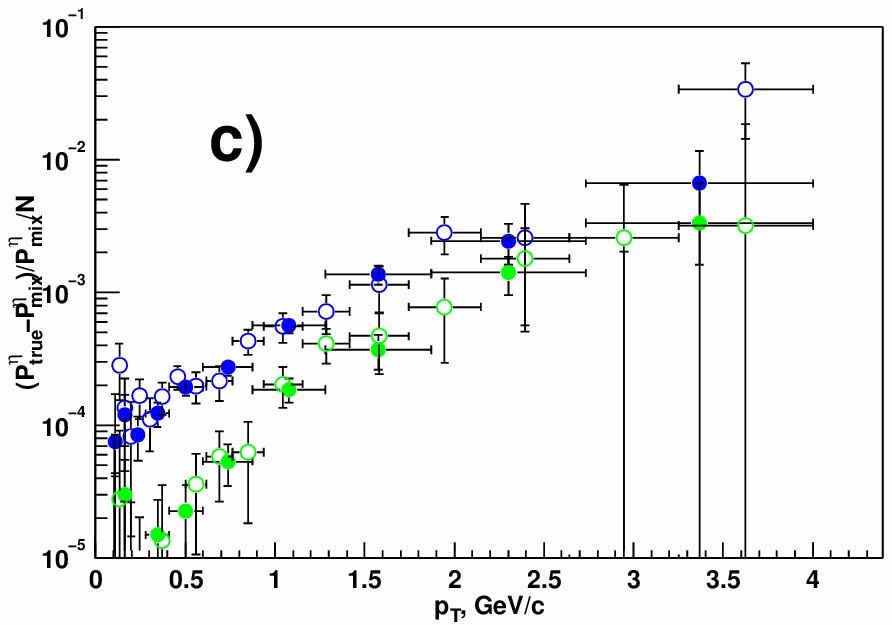}
\label{HIJING}
\end{minipage}
\end{figure}
\begin{figure}[htb]
\hspace{\fill}
\begin{minipage}[t]{70mm}
\epsfxsize=7cm
\centerline{\epsfbox{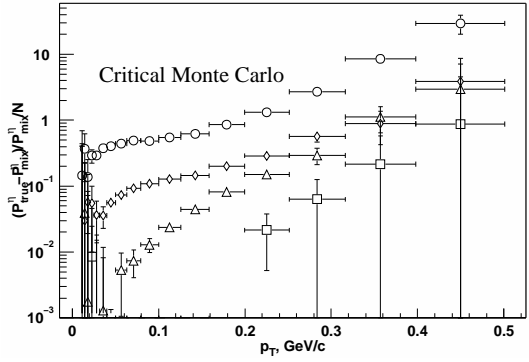}}
\end{minipage}
\hspace{\fill}
\begin{minipage}[b]{40mm}
\caption{\
$(P^{\eta}_{true}-P^{\eta}_{mix})/P^{\eta}_{mix}/N$ from the
Critical MC generator. 
Events with 20 to 30 charged tracks in the STAR
acceptance are analyzed.
$\bigcirc$ -- scale 1, $\Diamond$ -- scale 2, $\bigtriangleup$ -- scale 3,
$\Box$ -- scale 4.
}
\label{CMC_eta}
\end{minipage}
\end{figure}
 Figure  \ref{HIJING}
demonstrates observability of the HIJING dynamic effects in
our analysis.
\footnote{
In all MC generators, no GEANT and no response simulation is done.
Instead, only stable charged particles ($e$,$\mu$,$\pi$,$K$,$p$) and their
antiparticles from the generator output are considered, provided that
they fit into the STAR TPC fiducial $\eta$ acceptance $|\eta| \le 1$.
Momentum resolution and $p_T$ acceptance are not simulated.
}
We  see that, first, the difference between the true
and mixed  events is noticeable  and can be  studied as a  function of
$p_T$ with  the present HIJING  statistics of around  $1.6\times 10^5$
events.  Second, the open and closed symbols, which correspond to
different $p_T$ bin sizes, appear to fall  on the same curve after the $1/N$
normalization, where  $N$ is a $p_T$  bin multiplicity, as would  be the
case for ``coherent'' (see Section\ref{coherence}) $p_T$ bins.  
Third, the rise of the signal with $p_T$ is due to the fact
that high $p_T$ is dominated by jet production.
As far as the $p_T$ ``coherence''
 is concerned, one would expect that a high
$p_T$ parton, creating hadrons via fragmentation, produces similar 
($\eta$,$\phi$) patterns at different $p_T$ as the energy sharing
among the secondaries proceeds, and thus the coherent interference
of $p_T$ patterns is natural for 
this mechanism of particle production.
These signals in HIJING are gone when jet production is turned off in the
generator.
Ability to study jet textures at soft, as well as high, $p_T$ means 
that the study promises to be very informative because majority
of the reconstructed tracks will be utilized.

CMC  is  Critical Monte  Carlo  generator  created  by N.Antoniou  and
coworkers \cite{CMC}. 
In the framework of an effective action approach,
these authors simulate a system 
 undergoing a second order QCD phase transition.
The $\eta$ signal at low $p_T$ (Fig. \ref{CMC_eta}) 
is much stronger than seen in HIJING and
is dominated by the coarse scale.

\section{STAR measurements of dynamic textures}
\label{STAR_data}
Elliptic flow is a prominent large scale dynamic texture effect
already well measured at RHIC\cite{v2_RHIC}.
The DWT approach localizes elliptic flow on scales 2 and, to some degree, 3
of the azimuthal observables.
In this report, we ignore flow and concentrate on the $\eta$
observables.
\begin{figure}
\epsfxsize=10.5cm
\epsfbox{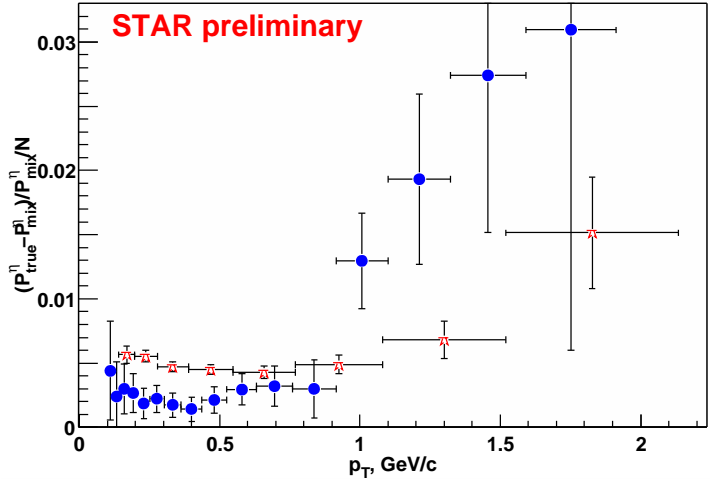}
\caption{\label{STAR_peripheral}
$(P^{\eta}_{true}-P^{\eta}_{mix})/P^{\eta}_{mix}/N$ for scale 1,
peripheral events. Open stars -- STAR data
for $\sqrt{s}=200$ GeV, $0.014<\mbox{mult}/n_0<0.1$.
\textcolor{blue}{$\bullet$} -- HIJING at the same energy, 
$\mbox{mult}/n_0<0.1$.
}
\end{figure}
\begin{figure}[htb]
\epsfxsize=10.5cm
\epsfbox{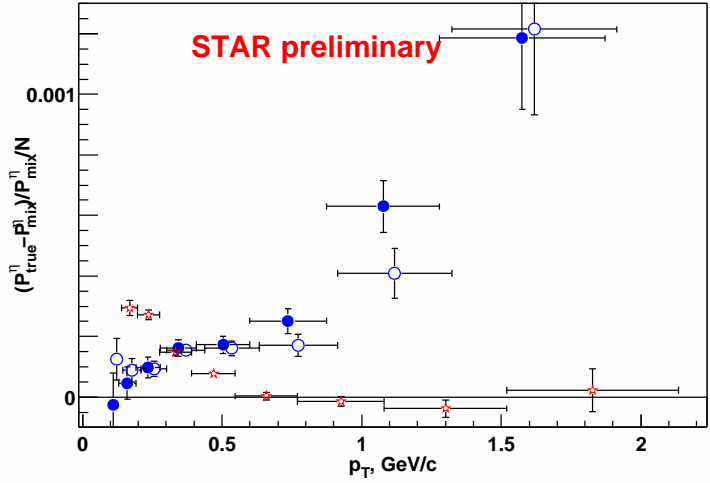}
\caption{$(P^{\eta}_{true}-P^{\eta}_{mix})/P^{\eta}_{mix}/N$ for scale 1,
central events ($0.65<\mbox{mult}/n_0<1.$).
 Open stars  -- STAR data for $\sqrt{s}=200$ GeV.
\textcolor{blue}{$\bullet$} -- regular HIJING;
\textcolor{blue}{$\circ$} -- HIJING with jet quenching, 
both at $\sqrt{s}=130 GeV$.
}
\label{STAR_central}
\end{figure}

Fig. \ref{STAR_peripheral} presents the STAR measurements of long range 
(scale 1) fluctuations in peripheral ($0.014<\mbox{mult}/n_0<0.1$) collisions
and compares them with HIJING simulations.
Qualitatively, both sets of points behave similarly:  a region
of nearly flat or falling behavior around mean $p_T$
is replaced by a rising trend for $p_T > 0.8$ GeV/c.
This trend  has already been discussed in Section
\ref{jets_in_HIJING} and is due to jets.
The HIJING signal is below the STAR data  at low $p_T$, but
reaches higher values at higher $p_T$; its rise with $p_T$ is stronger.
From this figure we conclude that the fluctuations in
local hadron density due to jet production are observable at RHIC in the soft
$p_T$ range ($p_T<2$ GeV), and that their \emph{qualitative} features
are reasonably well described by
a super-position of independent nucleon-nucleon collisions based on the
physics learned from $pp(\bar{p})$ and $e^+e^-$ 
experiments at comparable energies.
Quantitatively speaking,
we keep in mind that due to nuclear shadowing effect
\cite{nuclear_shadowing}, peripheral Au+Au
events are not supposed to be identical to elementary collisions.
A comparison of $pp$, dAu and AuAu data from RHIC will shed more light on
this effect.
In the absence of experimental data on nuclear shadowing of gluons,
HIJING assumes\cite{HIJING} equivalence of the effect for quarks and gluons.

Next look at a central sample (Fig. \ref{STAR_central}) --
there is a remarkable difference: we now see a change in the $p_T$ trend above
$p_T=0.6$ GeV.
Instead of rising with $p_T$ (as in the peripheral events), 
the STAR data points become consistent with 0.
The $p_T$ trends in the data and HIJING look opposite: the model still
predicts a monotonic rise with $p_T$.
Can there be a single explanation to both disappearance of texture at moderate
$p_T$ and its enhancement at low $p_T$?
The hypothetical deconfined medium is expected to suppress jet production
via dissipative processes (jet quenching) 
\cite{deconf_energy_loss}.
The medium-induced energy loss per unit of length
is proportional to the size of the medium and thus, the effect grows 
non-linearly with system size.
Suppression of hadron yields at high $p_T$ in central AuAu events with respect
to scaled $pp$ and peripheral collisions 
has been reported\cite{high_pT_suppression}
and interpreted as an evidence of medium effects (possibly, nuclear shadowing
\cite{nuclear_shadowing}).
Jet quenching is modeled in HIJING,
and is seen (compare two sets of HIJING points in Fig.\ref{STAR_central})
to affect the texture observable somewhat. 
If the dissipation takes place, one may expect that
as jets and mini-jets thermalize, the textures associated with them
migrate towards mean $p_T$.
A transport model would be needed in order to simulate such a process.
However, the low pT fluctuations may have an independent origin,
unrelated directly to the partonic energy loss in medium.

\section{Conclusions}

A non-trivial picture of texture effects emerges when the 
DWT power spectrum technique is applied to AuAu data from RHIC.
Long range ($\Delta\eta \approx 1$)
pseudo-rapidity fluctuations at soft $p_T$ are observed 
in peripheral events and identified with jets and mini-jets.
In central events, these fluctuations are not seen, which indicates
a change in the properties of the medium.
Large scale of the effect points to its early origin.
An excess of fluctuations at low $p_T$ compared to HIJING is seen
in peripheral and central events.

\section{Acknowledgment}
I am grateful to Nikos Antoniou and Fotis Diakonos for providing me with
simulated phase transition events to establish the sensitivity of the technique
to critical phenomena.

\end{document}